# A graph-based knowledge representation and pattern mining supporting the Digital Twin creation of existing manufacturing systems

Dominik Braun
Graduate School of Excellence
advanced Manufacturing Engineering
(GSaME)
University of Stutgart
Stuttgart, Germany
dominik.braun@gsame.uni-stuttgart.de

Timo Müller
Institute of Industrial Automation and
Software Engineering
University of Stuttgart
Stuttgart, Germany
timo.mueller@ias.uni-stuttgart.de

Nada Sahlab
Institute of Industrial Automation and
Software Engineering
University of Stuttgart
Stuttgart, Germany
nada.sahlab@ias.uni-stuttgart.de

Nasser Jazdi
Institute of Industrial Automation and
Software Engineering
University of Stuttgart
Stuttgart, Germany
nasser.jazdi@ias.uni-stuttgart.de

Wolfgang Schlögl
Digital Engineering
Siemens AG
Nuremberg, Germany
schloegl.wolfgang@siemens.com

Michael Weyrich
Institute of Industrial Automation and
Software Engineering
University of Stuttgart
Stuttgart, Germany
michael.weyrich@ias.uni-stuttgart.de

*Abstract*—The creation of a Digital Twin for existing manufacturing systems, so-called brownfield systems, is a challenging task due to the needed expert knowledge about the structure of brownfield systems and the effort to realize the digital models. Several approaches and methods have already been proposed that at least partially digitalize the information about a brownfield manufacturing system. A Digital Twin requires linked information from multiple sources. This paper presents a graph-based approach to merge information from heterogeneous sources. Furthermore, the approach provides a way to automatically identify templates using graph structure analysis to facilitate further work with the resulting Digital Twin and its further enhancement.

## I. Introduction

The customers' increasing demand for individualized products, shorter innovation cycles, as well as the increased economic volatility, lead to increased adaptions of existing manufacturing systems [1]. The industrial 4.0 trend addressed this and similar challenges by introducing a digital replica to the real plant, the so-called Digital Twin. The Digital Twin as a cyber construct and the real manufacturing systems as a physical counterpart form a cyber-physical production system [2]. Enhancing existing manufacturing systems, so-called brownfield systems, with a Digital Twin to form cyber-physical systems is a challenging and error-prone task [3]. It requires high manual effort and expertise to reverse-engineer models and relations of the brownfield systems [4]. Therefore, several publications e.g. [4, 5] deal with the model creation, synchronization, or reverse-engineering of a plant. They support the digitalization of single aspects of the Digital Twin. At least as important as the individual models are the relations between them. However, there are almost no methods that can reverse engineer these relations. Therefore, a methodology based on the control code, the material position data, and the IO signals was proposed to reverse-engineer these relations from brownfield systems and support their digitalization [6]. The information retrieval from different types of information sources was presented in previous publications [6, 7] and needs to be merged in the next step. This publication will briefly summarize the overall methodology and the information retrieval in section II. Afterward, the information merging from these sources is presented in section III. In this context, the task of merging information, the challenges, and the existing solutions are considered based on which, the own approach for merging the information about the relations is presented. Section IV discusses the benefits of the graph-based approach for merging information about the relations. Finally, section V describes the future research in this field and the applicability of the results in further work.

## II. Methodology for the Digital Twin creation of existing manufacturing systems

The proposed methodology aims to support the creation of the Digital Twin of existing manufacturing systems by automating the relation creation between the models as the base of the Digital Twin. According to Ashtari *et al.* [8], a Digital Twin consists of a digital replica with three characteristics: (1) it is simulatable, (2) a digital replica needs to be synchronized, and (3) active data acquisition. The digital replica as the base of the Digital Twin consists of models and the relations between them (see Fig. 1).

Several publications deal with single model creation such as [4, 9, 10]. To create a comprehensive Digital Twin of an existing manufacturing system, the relations are at least as important as the models and difficult to obtain as inter-domain experience and tool knowledge are needed. So far, there is only one methodology that is useable to obtain these relations [5]. But the presented anchor point methodology in [5] is only suitable for new production systems as a naming convention needs to be applied to the PLC code during engineering reflecting the relations. Therefore, a methodology for an automated, digital reproduction of the relations of an existing manufacturing system as a base of the Digital Twin was presented in [6]. It consists of three information sources: (1) the PLC code and rule-based analytics [7], (2) position data and (3) IO data in a data-based analytic approach [6]. These information sources form the needed information base to represent the relations within a Digital Twin. Fig. 1 illustrates the information sources, analysis methods, and the connection to the knowledge base. The topic of this paper deals with the information merging and storing inside the gain knowledge representation as well as the pattern mining from it.

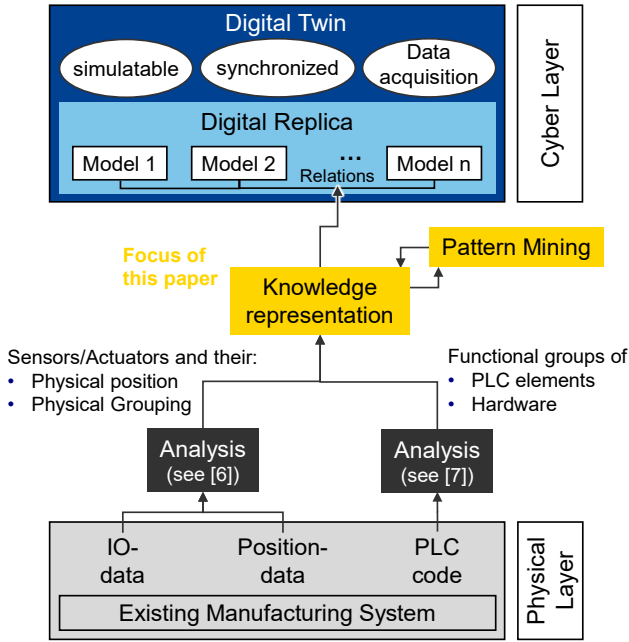

Fig. 1. The methodology (according to [6]) for an automated creation of the Digital Twin (according to [8]) of brownfield production systems

In the overall methodology, the resulting knowledge base can finally be mapped in a project-specific environment and result in an implementation of the Digital Twin. This methodology creates the relation structure of the models and thus needs to be enriched with further information, e.g. with an additional, detailed CAD model in the digital replica, to be usable for specific applications. Therefore, information from other sources depending on the intended use case must be added either to the combined knowledge base resulting from this research or to the final Digital Twin implementation.

III. KNOWLEDGE REPRESENTATION AND PATTERN MINING

This paper deals with the various information gained from three information sources and how they can be merged into a shared storage. The following sub-sections describe the tasks that must be completed to achieve a joined representation and the challenges to fulfill these tasks. Subsequently, existing solutions are discussed and an own approach based on the existing solutions is presented.

*A. Task: derivation of a knowledge representation for repetitive pattern mining*

The three data sources presented in section II and Fig. 1 contain different information about the relations of existing manufacturing systems. This is retrieved with various methodologies as described in [6, 7]. The task is to connect information from all three sources in Fig. 1 to a coherent information structure. Besides that, it is recognized that the information gained from these sources about a production system contains repeating patterns due to reused assemblies and components in a system. Therefore, the information merging and storing approach must support pattern searching in the resulting information representation to identify repetitive patterns of elements and relations. These repetitive structures represent hardware components or assemblies and software functions of production systems which are used several times in the production system. Identifying these recurring structures enables a reduction of redundant data by using templates for the repetitive elements according to the class instantiation in object-oriented programming. A less redundant and smaller dataset simplifies data handling and later works with the resulting Digital Twin for its users due to the dataset size. Besides that, changes to the Digital Twin models are faster to deploy, because changes to a component group only need to be applied once for the template and not several times for each occurrence.

*B. Challenges regarding the information storage and analysis of existing manufacturing systems*

The information generated from the three sources in Fig. 1 and need to be stored in the shared storage is about (1) the functional components and their grouping as a functional module, and (2) the physical arrangement of the components and their groups. This information is characterized by multiple elements, information describing these elements, and many relations between them. These elements are also called nodes and form together with the relations a network such as a tree or a meshed network structure. The nodes themselves contain only a part of the information. The relevant information is in contrast to conventional data storage not contained in the element itself but in the relations between the single elements. Furthermore, to process the vast data and find repetitive groups of element arrangements without them being pre-defined is another challenge.

*C. Existing solutions for knowledge storage and pattern mining algorithms*

There are several frequently used data storage solutions and a common one is a SQL database for relational data. Furthermore, specialized databases for time-series data, documents, key-value pairs, or graph databases evolved. Besides that, there exist various data formats such as XML, JSON, or CSV. These file-based formats are mostly used as exchange formats rather than persistent storage for a large amount of data. This is because managed database systems have a better performance with large amounts of data, parallel data access, and querying data is more convenient than with a file-based data storage [11]. Data with a high variety in the data structure, with many connections between entries or large time-series of data points, cannot be stored efficiently in relational databases either, or require extensive data preparation and transformation processes. Therefore, time-series, document, or graph databases as specialized NoSQL databases evolved. In [12] the most common NoSQL databases are compared and the strength and weaknesses are discussed. For the storage of the data gained in the methodology, a graph database is the most promising approach, since it is well suited to handle intensively connected data and traverse multiple relations during data processing.

Neither of the presented data storage solutions contains a built-in function to identify repetitive structures. But there exist several publications about frequent subgraph mining in the field of graph databases and numerous algorithms are developed to identify repetitive structures in graphs. The authors in [13] compared some common algorithms and their performance in a structured way based on several graph datasets for molecular data. The gSpan algorithm [14] performs well for data with small or medium subgraph sizes and requires less computational resources because it does not use embedding lists. Algorithms with embedding lists e.g. FFSM [15] work faster for datasets with large subgraphs.

Because it is expected to mainly find small repetitive structures in a production plant, the gSpan algorithm will be focused on in the research. Besides that, there are several specialized gSpan extensions such as for closed subgraph mining [16] or the subgraph mining in conceptual graphs [17].

### D. Proposed approach to merge information in knowledge representations and identify repeating patterns

The overall methodology deals with plenty of nodes and edges between them extracted from the three information sources. For this reason, a graph database is the best solution to store information and will be used to store the incoming information. It is optimized for strongly related data and enables effective querying. Additional to a graph database, an underlying ontology will be used as presented in [18] to increase the comprehensibility of the stored information. This hybrid hierarchy-based context tier model uses a labeled property graph for the implementation of the ABox representing the knowledge about the system. The model divides the nodes and edges containing information about the system into four levels of abstraction from the domain internal view, the inter-domain view over the systems-of-systems perspective to the entire scope comprising the systems environment. This research about the reverse engineering of the Digital Twin relations focuses on the two lower tiers of the domain internal and inter-domain contextual relations according to [18]. The underlying ontologies are a stack, of a generic basic ontology using ontology design patterns [19] for mechatronic systems and Digital Twins to increase the semantic and reusability for other manufacturing systems. On top of the basic ontology is the terminology box (TBox) and refines it for the system-specific characteristics (see Fig. 2). In this research, a specific ontology for the FlexCell project in the Arena2036 (Active Research Environment for the Next Generation of Automobile) is used to evaluate the approach based on an industrial research project system.

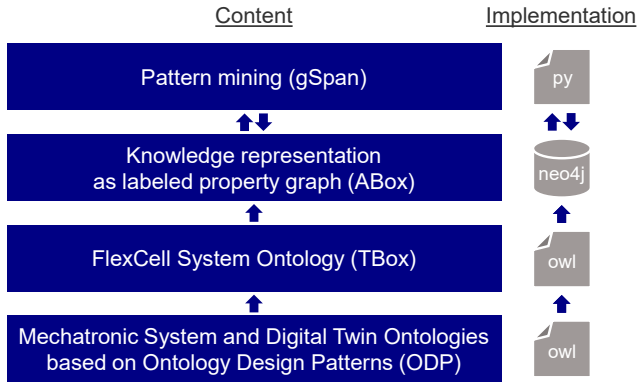

Fig. 2. Knowledge representation and pattern mining based on the hybrid hierarchy-based tier model

The information from the three sources is stored in the ABox. The graph connection to each source is done on a multi-level procedure. On the one hand, the node labels, mainly the name and the type label, are used to find equal nodes. Because the three paths contain elements with names from the PLC code either directly (PLC as data source, IO signals from the PLC through OPC UA) or indirectly (position data combined with the IO signals) the node name labels are comparable and can be used for the merging process. On the other hand, ontologies are used to identify semantically identical structures and merge them. This combination creates one large connected graph describing the system structure.

Consequently, the cgSpan algorithm [17] is promising for the implementation outlined above as it also takes into account the contextual knowledge of the ontology for the subgraph identification. Besides that, the frequent subgraph mining does not have to be a specialized implementation excluding unclosed subgraphs because the small patterns are important as well as they describe the smallest repeating system component groups. These component groups should be identified and marked as reusable templates as well and can be combined into larger templates representing a complete subgraph afterward. With this approach, high flexibility regarding structural changes to single elements will be established.

## IV. DISCUSSION ON ADDED VALUE

One benefit of the information merging and storing approach using the hybrid hierarchy-based context tier model is the high flexibility and reusability for several other production systems with low effort. If not needed for high expressiveness, the system-specific upper ontology (TBox) can be reduced or even omitted and the graph-based information storage is directly based on the basic ontology built from the ontology design patterns. Generally, the graph-based information storage is well suited for representing and retrieving information about the Digital Twin relations with many heavily related nodes and the efficient querying possibilities along with several levels of relations.

An essential benefit of the proposed approach is the available frequent subgraph mining algorithms for graph-based implementations with their specialized extensions for conceptual graphs. Fig. 3 shows the physical FlexCell system and prototypical implementation of its ABox, which consists of 242 nodes and 402 relations. The research focuses on the lower two tiers of the ABox and excludes the green and red nodes of the ABox but still, there are a vast of tier one and two nodes respectively edges (blue/yellow nodes/edges). Through the identification of repetitive subgraphs by applying frequent subgraph mining to the ABox, the information load becomes manageable for an engineer. The extract from the ABox in the lower-left corner of Fig. 3 contains the nodes of the lower level of the warehouse of the FlexCell system (depicted in the foreground of the FlexCell system image). With the subgraph mining, the four instances of the storage row pattern are identified (each instance has a different color, see Fig. 3).

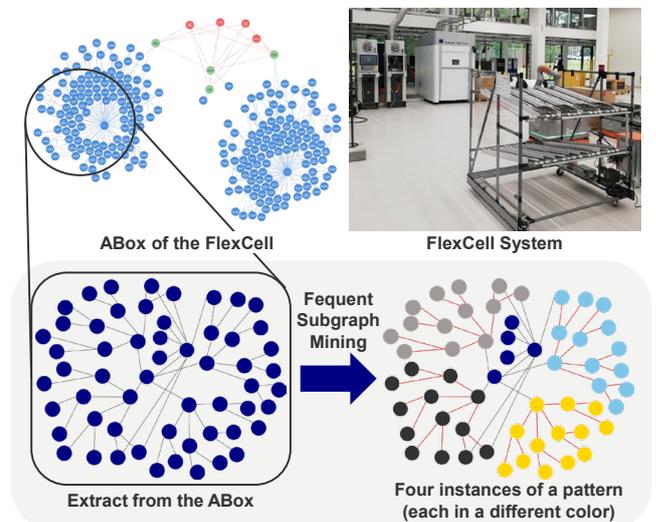

Fig. 3. Identification of the four times repeating subgraph pattern

The resulting graph with the identification of the repetitive subgraphs is firstly easier to handle for engineers because it is better structured and secondly, the implementation as digital models is modular and easy to reuse as well as modifiable due to the manageable smaller templates.

## V. Conclusion and Outlook

The reverse engineering of the relations of a brownfield system as the basis for the creation of a Digital Twin is a complex and time-extensive task. The proposed methodology for the automated extraction of these relations from brownfield systems and the implementation as a Digital Twin depends on three information sources. The information obtained must be merged and processed in a (re)usable and controllable manner. This work in progress presents the tasks that need to be fulfilled, their challenges, and discusses the existing realization options. Based on this, an own approach is presented, which is highlighted by:

- reusable knowledge representation for manufacturing systems based on the hybrid hierarchically tier model,
- merge the information from the information sources,
- repetitive pattern mining of often used mechatronic component groups based on frequent subgraph mining,
- a simpler structured knowledge representation based on the identified templated,
- a collective modification and easier reuse based on the identified templates,
- easier reuse of the knowledge for engineers due to the reduced vast of information.

In further research, some algorithms for frequent subgraph mining will be implemented and tested on the available information about brownfield manufacturing systems from the three information sources. The influence of the ontologies on the frequent subgraph mining will be examined in more detail and evaluated based on the intelligent warehouse in the FlexCell project.

## Acknowledgment

This work was supported by the Deutsche Forschungs-gemeinschaft DFG (German Research Foundation) within the Exzellenzinitiative (Excellence Initiative) – GSC 262 and the Landesministerium für Wissenschaft, Forschung und Kunst Baden-Württemberg (Ministry of Science, Research and the Arts of the State of Baden-Wurttemberg) within the Nachhaltigkeitsförderung (sustainability support) of the projects of the Exzellenzinitiative II.